\documentclass[a4paper]{article}

\usepackage[margin=1in]{geometry} 

\usepackage{amsmath}
\usepackage{amsthm}
\usepackage{amssymb}

\usepackage{booktabs}

\usepackage[utf8]{inputenc}
\usepackage{hyperref}
\hypersetup{
	unicode,
	pdfauthor={James Bono},
	pdftitle={Randomized Controlled Trials for Phish Triage Agent},
	pdfproducer={LaTeX},
	pdfcreator={pdflatex}
}

\usepackage{natbib}
\setcitestyle{authoryear,open={(},close={)}} 

\theoremstyle{plain}

\theoremstyle{definition}

\usepackage{graphicx, color}
\graphicspath{{fig/}}
\usepackage{subcaption}
\usepackage{xcolor}

\usepackage{algorithm, algpseudocode} 
\usepackage{mathrsfs} 
\usepackage{multirow}

\usepackage{lipsum}

\usepackage{tablefootnote}

\title{Randomized Controlled Trials for Conditional Access Optimization Agent}
\author{James Bono\thanks{james.bono@microsoft.com; The authors would like to thank the following contributors: Nikhil Reddy Boreddy, Michael Browning, Jordan Dahl, Justin Grana, Katerina Marazopoulou, Mitch Muro, Rahul Prakash, and Tracy Zhen} \and Beibei Cheng \and Joaquin Lozano }

\date{  
	Microsoft Corporation \\
	October 2025
}

\begin{document}
	\maketitle
	
	\begin{abstract}
    AI agents are increasingly deployed to automate complex enterprise workflows, yet evidence of their effectiveness in identity governance is limited. We report results from the first randomized controlled trial (RCT) evaluating an AI agent for Conditional Access (CA) policy management in Microsoft Entra. The agent assists with four high-value tasks: policy merging, Zero-Trust baseline gap detection, phased rollout planning, and user-policy alignment. In a production-grade environment, 162 identity administrators were randomly assigned to a control group (no agent) or treatment group (agent-assisted) and asked to perform these tasks. Agent access produced substantial gains: accuracy improved by 48\% and task completion time decreased by 43\% while holding accuracy constant. The largest benefits emerged on cognitively demanding tasks such as baseline gap detection. These findings demonstrate that purpose-built AI agents can significantly enhance both speed and accuracy in identity administration.

	\end{abstract}

	
\section{Introduction}
\label{sec:introduction}

AI agents are increasingly deployed not just as tools but as collaborators in economic production. The administrative backbone of the digital economy -- IT infrastructure -- offers a fertile domain to evaluate AI's impact. Identity management in particular sits at the junction of security and productivity: its effectiveness determines both organizational safety and the efficiency of digital workflows. The question of whether purpose-built AI agents can meaningfully augment this kind of administrative work is therefore central not only to IT operations but also to the broader debate about how AI reshapes labor and productivity.

The ability to enforce finely grained conditional access (CA) policies is critical to identity security. Yet crafting, tuning, and auditing these policies is cognitively and operationally expensive. In this paper, we present the first randomized controlled trial (RCT) to evaluate whether an AI agent that suggests or optimizes CA rules -- the Entra Conditional Access Optimization Agent (``the agent'') -- can materially boost the productivity of identity administrators (``admins''). We randomize a diverse set of admins into Treatment and Control groups, granting the former access to the agent’s suggestions while the latter perform tasks unaided. Our results show robust gains in both task speed and correctness: administrators assisted by the agent complete tasks 43\% faster while holding accuracy constant and are 48\% more accurate. These findings open a path toward AI-augmented identity governance, highlighting heterogeneity of gains across task types.

The agent is designed to perform a set of four complex and labor-intensive tasks associated with CA policy management. These include i) merging existing policies to reduce policy overlap; ii) identifying gaps in Zero-Trust baseline policies and building policies to fill them; iii) designing phased rollout of new policies to ensure smooth deployment and minimize operational impacts; and iv) ensuring newly added users are included in appropriate existing policies to achieve Zero-Trust coverage. The agent can be configured to make suggestions only or to autonomously create new policies in report-only mode for an admin to review and approve. 


We recruit a targeted population of 162 identity admins to perform the four CA policy management tasks in a purpose-built, production-grade environment. Subjects interact with the environment and record their responses to the tasks via an instrumented survey platform. Subjects are split into two groups: a Control group without access to the agent and a Treatment group that has access to the agent's suggestions.

\paragraph{Research Questions.} Our research questions are motivated by what a CIO would want to know prior to adoption:
\begin{itemize}
  \item \textbf{RQ1 (Accuracy):} Does using the agent improve the quality of admin decisions?
  \item \textbf{RQ2 (Speed):} Holding accuracy constant, how much time does the agent save an admin?
\end{itemize}

AI’s measured productivity effects are now well-documented across writing, customer support, and software development, with sizable average gains but marked heterogeneity by task and worker experience \citep{Noy2023,Brynjolfsson2023,Peng2023,Cui2024}. Within enterprise IT operations, evidence likewise points to faster, more accurate work from generative assistants \citep{BonoXu2024,Edelman2024} and associated improvements in live operational metrics \citep{Bono2024,Bono2025}. Moving from copilots to agents, some expect planning, memory, and tool use expand the scope for automation and complementarity \citep{Korinek2025}. Agent experiments, like the present one, have begun to quantify agent effects in collaborative and technical workflows, highlighting design, trust, and integration frictions \citep{Ju2025,Chen2025,Becker2025}. However, from a macroeconomic perspective, rapid adoption coexists with tempered forecasts for aggregate total factor productivity gains absent complementary organizational change \citep{Bick2024,Acemoglu2024,Calvino2025}. Against this backdrop, our contribution is a field-grade RCT in identity governance that measures how a purpose-built agent shifts speed and accuracy for professional administrators.

This paper is organized as follows. First, we provide a detailed description of our experimental methods. Then, we present the results of our analysis and address the research questions above. Finally, we discuss the implications of these findings and future work.

\section{Methodology}
\label{sec:methodology}

We follow a standard RCT protocol to identify and estimate the causal effect of the agent on accuracy and task completion time. That is, we randomly assign half of our subjects to the Control group, which only allows them use of the Entra admin center without the agent. The other half are assigned to the Treatment group, which gives them the additional ability to use the agent's output when responding to the experimental tasks. To make our design as realistic as possible, we placed no restrictions on how subjects used other tools, such as other AI tools or web searches. Therefore, our measurements reflect the incremental benefit of the agent on top of the admin centers and other tools like web searches and other AI.

We recruited our subject pool through Upwork, a marketplace for freelancers. We required subjects to have some experience in identity management, to be proficient in reading and writing English, and to have a positive reputation on Upwork. We told them to expect the tasks to take up to 1.5 hours to complete. We offered performance incentives for combined speed and accuracy. Specifically, we multiplied the percentiles of each subject’s speed and accuracy and awarded payments on the following scale:
\begin{itemize}
    \item Top 10\% in combined speed and accuracy: \$85
    \item Top 10-20\% in combined speed and accuracy: \$70
    \item Top 20-30\% in combined speed and accuracy: \$55
    \item Completed in good faith but did not reach the top 30\% in combined speed and accuracy: \$35
    \item Show up fee: \$15
\end{itemize}

We had 162 subjects complete the task. Our subject pool embodied a high level of experience across identity management. Table \ref{tab:experience-control-treatment} presents a breakdown of subjects' self-reported experience level by experimental group. We note that the groups are not perfectly balanced by experience level. However, when including experience level in regressions to measure speed and accuracy differences, we only find a statistically significant effect for the accuracy metric for one of our tasks -- Policy Merge. Therefore, our results on Policy Merge accuracy control for experience level. We do this for completeness only and note that the results are robust to this choice.

\begin{table}[htbp]
\centering
\caption{Experience Distribution Across Control and Treatment Groups}
\label{tab:experience-control-treatment}
\begin{tabular}{lccc}
\toprule
\textbf{Experience Level} & \textbf{Control} & \textbf{Treatment} & \textbf{Total} \\
\midrule
Less than a year           & 0  & 5  & 5   \\
Between 1 and 3 years      & 20 & 8  & 28  \\
More than 3 years          & 63 & 66 & 129 \\
Total                      & 83 & 79 & 162 \\
\bottomrule
\end{tabular}
\end{table}

We built our laboratory within Microsoft Entra, a cloud-based identity admin center. We simulated data in this environment to reflect what might be found in the environment of a small organization. The data involved a variety of users, security groups, and CA policies. Subjects were given unique identities with read access to the environment and were asked to log in. They were then given instructions on how to navigate Entra to find the pages with information relevant to the experimental tasks. The treatment subjects were additionally given instructions on how to access the agent's suggestions. We presented subjects with four tasks:

\paragraph{Policy Merge:} Suggest merging two similar policies to reduce the number of overlapping policies. Subjects must identify the policies to merge, the reason for merging, and the content of the merged policy. This task is designed such that it has only a single correct answer. Therefore, we employ the same production-grade evaluation to the responses that we use to evaluate the agent in live operations. The subject's score is zero if both policies are incorrect, 0.5 if one policy is correct, and one if both policies are correct.  

\paragraph{Missing Baselines:} Identify gaps in Zero-Trust baseline policies. Subjects are provided with a complete list of Zero-Trust baseline policies and must suggest which Zero-Trust baseline policies are missing in the tenant. The task is graded as a classification problem -- classifying policies as missing or not. Subject scores are their F1 scores. 

\paragraph{Phased Rollout:} Design a phased rollout plan for a missing Zero-Trust policy. Subjects must identify appropriate groups for each of the first four phases of a standard five-phase plan and provide justification. We use an LLM-based evaluation that compares the subject's response against the tenant ground truth and scoring guidelines tied to four categories: Validity (0.35 pts), phasing strategy quality (0.40 pts), safety and rollback (0.05 pts), and rationale (0.20 pts). The full evaluation is provided in appendix \ref{sec:phasedrollouteval}.

\paragraph{Add New Users:} Add new users to existing policies. Subjects must identify which new users are not in scope of existing Zero-Trust policies and add them to ensure coverage. This task is designed such that it has only a single correct answer. Subjects who identify the correct policy and at least half of the correct users receive a score of one. All others receive a score of zero. 
\vspace{1em}

In measuring the effect of the agent on task completion time, we note that subjects will judge whether spending more time on a task is likely to be rewarded with a higher score. \emph{Ex ante}, we believe this judgment is likely to be different across experimental groups, e.g., Control group users may judge a task too difficult and give up quickly, whereas the Treatment group's use of the agent may reward additional time spent. Hence, to ensure we are making appropriate comparisons, we calculate differences in speed while holding accuracy constant. For Add New Users and Policy Merge tasks, we compare times for subjects that answered correctly (score of one). For Missing Baselines and Policy Merge tasks, we calculate the average additional time it would take Control group subjects to achieve the Treatment group accuracy. Details of this calculation are included in appendix \ref{sec:calculatetimesavings}.

\section{Results}
\label{sec:results}

\subsection{Accuracy}
The agent improves accuracy by 48\% across all four tasks. The results are reported in table \ref{tab:accuracy}. The biggest increase comes from Missing Baselines, where the agent's effect is 204\%. On this task, the Control group's F1 score is just 0.15, indicating that, without the agent, admins have difficulty assessing the tenant's Zero-Trust coverage. Although the Treatment group's F1 score is still a modest 0.46 on this difficult but important task, that score demonstrates significant improvement. 

\begin{table}[htbp]
\centering
\caption{Accuracy Gains Across Experimental Tasks}
\label{tab:accuracy}
\begin{tabular}{lccccc}
\toprule
\textbf{Task} & \textbf{Control Mean} & \textbf{Treatment Mean} & \textbf{Effect Size} & \textbf{\% Change} & \textbf{p-value} \\
\midrule
Policy Merge\tablefootnote{The effect size, percentage change, and p-value for Policy Merge were all calculated from a regression including subjects' experience levels. This is the only metric in our study where experience levels had statistically significant effects.}      & 0.44 & 0.60 & 0.37 & 51  & 0.028 \\
Missing Baselines & 0.15 & 0.46 & 0.84 & 204 & $<$0.001 \\
Phased Rollout    & 0.72 & 0.85 & 0.50 & 18  & 0.002 \\
Add Users         & 0.49 & 0.76 & 0.57 & 54  & $<$0.001 \\
Total Score       & 1.80 & 2.66 & 0.93 & 48  & $<$0.001 \\
\bottomrule
\end{tabular}
\end{table}

Phased Rollout shows the smallest effect in percentage terms. However, this was the only task that did not use an F1 score but rather used an LLM-based evaluation. Hence, the scale is not standardized in the same way as the F1 scores. This is why we also report Cohen's d for the ``Effect Size'' in table \ref{tab:accuracy}. With this metric, the improvement to Phased Rollout is 0.50 standard deviations, which is neither the lowest nor the highest. Indeed, the effect size for Policy Merge is the smallest. That may be an artifact of the relatively small number of policies in the experimental tenant, making it relatively easy for Control subjects to identify overlap.

\subsection{Speed}

\begin{figure}[htbp]
    \centering
    \caption{Total Time Savings Holding Accuracy Constant}
    \includegraphics[width=1\linewidth]{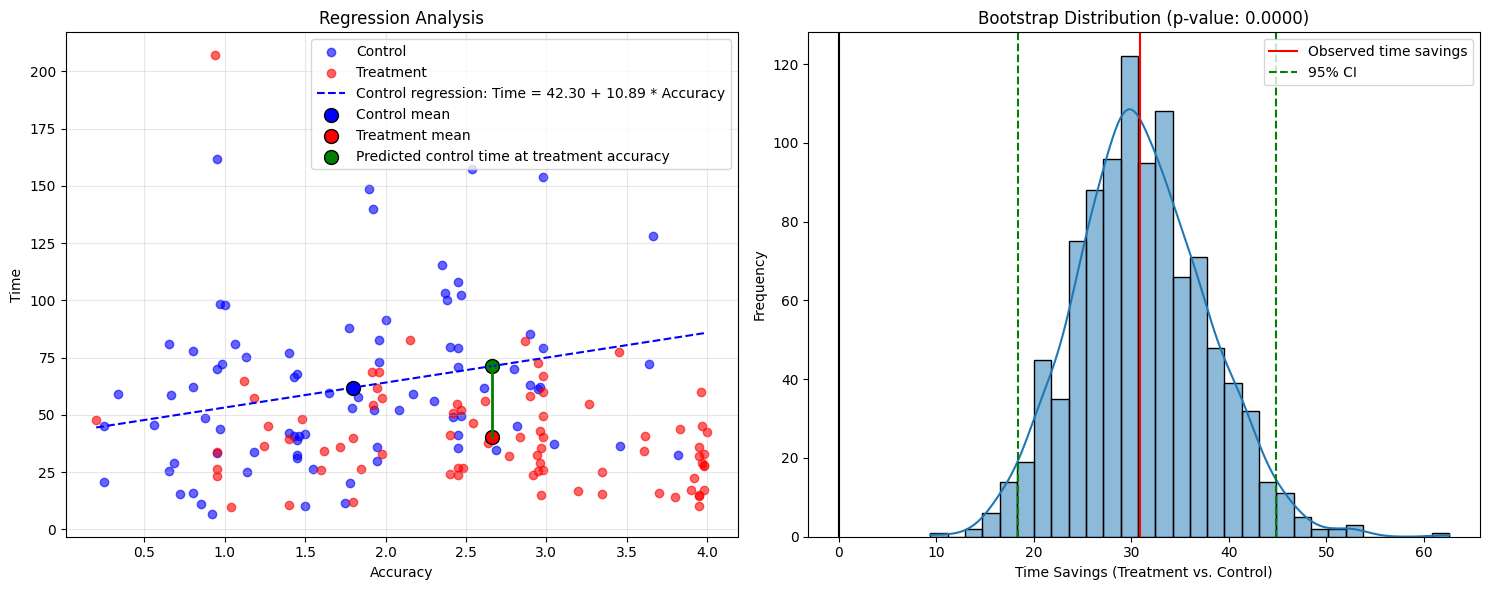}
    \label{fig:totalspeed}
\end{figure}

Holding accuracy constant, the Treatment group experiences a 43\% total time savings across all four tasks. The analysis for total time is shown in \ref{fig:totalspeed}. The left panel shows time as a function of accuracy for the Control group (blue line). The difference between the green dot and the red dot is the predicted time savings for the Treatment group. The right panel shows the bootstrap distribution with 95\% confidence intervals, where zero falls far outside the distribution. 

Table \ref{tab:time-accuracy-comparison} reports the analysis across all tasks. The time savings, holding accuracy constant, are large and statistically significant for every task. The asterisks (*) for Policy Merge and Add Users indicate times to get correct answers, as these binary scores require a different analysis than the continuous scores. 

\begin{table}[htbp]
\centering
\caption{Time and Accuracy Comparison Between Control and Treatment}
\label{tab:time-accuracy-comparison}
\begin{tabular}{lcccccc}
\toprule
\textbf{Task} & \textbf{Control} & \textbf{Control} & \textbf{Treatment} & \textbf{Treatment} &  \textbf{Time Savings} & \textbf{p-value} \\
& \textbf{Time} & \textbf{Accuracy} & \textbf{Time} & \textbf{Accuracy} & \textbf{(\%)} & \\
\midrule
Policy Merge      & 32.58$^{*}$ & 0.43 & 13.02$^{*}$ & 0.60 & 60 & 0.001 \\
Missing Baselines & 4.46        & 0.15 & 3.29        & 0.45 & 31 & 0.007 \\
Phased Rollout    & 16.42       & 0.72 & 11.39       & 0.85 & 37 & $<$0.001 \\
Add Users         & 16.60$^{*}$ & 0.49 & 9.12$^{*}$ & 0.76 & 45 & 0.001 \\
Total             & 61.86       & 1.80 & 40.40       & 2.66 & 43 & $<$0.001 \\
\bottomrule
\end{tabular}
\end{table}

\section{Discussion}
\label{sec:discussion}
Our randomized controlled trial provides clear evidence that a purpose-built AI agent can substantially improve both the speed and accuracy of identity admins performing complex CA policy management tasks. These findings contribute to the growing literature on AI’s productivity effects while extending it into an important administrative domain that has received little empirical attention.

The magnitude of the observed gains is notable. Accuracy improvements of 48\% on average, and more than 200\% for the Missing Baselines task, suggest that the agent meaningfully augments human judgment in areas where cognitive load and domain complexity are high. The Missing Baselines task, which requires holistic reasoning about Zero-Trust coverage, appears particularly well-suited to algorithmic assistance. By contrast, Policy Merge shows smaller effect sizes once differences in scoring scales are accounted for. This may be an artifact of the relatively small number of policies in the tenant, making it relatively easy for the Control group to find overlap.

Speed effects are similarly striking. Holding accuracy constant, the Treatment group completed tasks 43\% faster overall, with time savings of 60\% for Policy Merge. These results underscore the complementarity between human expertise and AI-generated suggestions: admins appear to leverage the agent to bypass time-intensive search and synthesis steps, reallocating effort toward validation and decision-making.

Several limitations warrant caution. First, while our experimental environment was production-grade, it remained a simulation; real-world deployments may introduce additional frictions such as organizational norms, compliance constraints, and integration costs. Second, our subject pool, though comprised of experienced identity admins, was drawn from a freelance marketplace and may not fully represent in-house enterprise admins. However, this concern is mitigated by the lack of an experience effect across the majority of our outcomes. Finally, we did not measure long-term learning effects or potential over-reliance on the agent.

Future research should explore longitudinal impacts, organizational adoption dynamics, and the interplay between agent autonomy and human oversight. Comparative studies across domains -- such as network security or compliance -- could further illuminate where agent-based augmentation delivers the greatest value.
\newpage

\bibliography{refs}
\newpage

\appendix

\section{Phased Rollout Evaluation}
\label{sec:phasedrollouteval}

\begin{verbatim}
# Phased Rollout Evaluation Guide: Zero-Trust Baseline - Block Legacy Authentication

## Overview
You are evaluating an identity administrator's submission for a phased rollout plan of 
the Zero-Trust baseline policy: Block Legacy Authentication.

Identity admin best practices require five phases:
1. Initial rollout to a very small, low-risk group to validate policy impact and 
ensure no unexpected disruptions. This stage is designed for close monitoring and
rapid feedback.
2. Expand rollout to a few additional small, low-risk groups to further validate 
policy behavior and gather broader feedback while still limiting potential impact.
3. Rollout to small-to-medium operational and technical groups to further increase 
coverage while still maintaining manageable risk and oversight.
4. Significantly increase coverage by targeting larger business and operational 
groups, representing a substantial portion of the tenant. This stage is intended to
validate policy impact at scale before full deployment.
5. Final stage: extend policy to all remaining users in the tenant, completing the
rollout.

## Inputs

### Subject Submission
- PHASES: An ordered list of phases 1-4
- For each phase:
  - groups: Comma-separated group object IDs
- Note: Phase 5 = All Users is assumed and is NOT provided

### Tenant Ground Truth
- GROUND_TRUTH: JSON including:
  - Groups (id, displayName, membership type, number of members)

## Task Reminder
For each of the FIRST FOUR phases, subject must list included group object IDs.

Important Caveat: Including Groups A and B is optional. If they are included in phases 
1-4, they should be included appropriately. However, there should be no deductions if 
they are not included in phases 1-4. 

## Your Task
Grade the subject's plan against conditional access and change-management best 
practices for blocking legacy authentication. Consider correctness (valid groups, 
constraints), safety (blast radius control, exceptions), and operational discipline 
(telemetry, rollback).

## Evaluation Rubric (100 points)

### A. Validity (40 pts)
- 20 pts: All listed group IDs exist in ground truth
- 20 pts: No contradictions (e.g., same group placed in multiple earlier phases 
without rationale or circular logic)

### B. Phasing Strategy Quality (60 pts)
- 35 pts: Sensible ring-based progression (pilot/canary low-risk → moderate → 
high-impact) with controlled blast radius. Risk-based grouping (e.g., test 
accounts, IT, low-risk departments first; high-criticality or complex environments 
later)
- 25 pts: Each phase should represent a larger deployment increment than the last, 
e.g., stage 4 deploys to more users than stage 3  

## Deductions (Examples)
- -35 pts: Assigns high-impact/unknown-dependency groups in Phase 1 without 
justification
- -20 pts: Lists group IDs that do not exist in the ground truth

## Instructions to the Grader Model

1. Check structure and constraints first (phases 1-4 present, valid IDs)

2. Evaluate phasing strategy for sensible, risk-aware progression and blast-radius 
control

5. Score per rubric, applying deductions proportional to the violations (e.g., 
-5 pts if only one phase has group IDs that do not exist in the ground truth)

6. Output a JSON object: 'score' 0-100; 'section_scores' per rubric; 'feedback' 
3-6 concise bullets with specific, actionable notes

## Output Format

```json
{
  "score": <0-100>,
  "section_scores": {
    "validity": <0-40>,
    "phasing_strategy_quality": <0-60>,
  },
  "feedback": [
    "Bullet 1",
    "Bullet 2",
    "Bullet 3"
  ]
}
```

Note: Feedback should contain 3-6 concise bullets with specific, actionable notes.
\end{verbatim}

\section{Calculating Time Savings Holding Accuracy Constant}
\label{sec:calculatetimesavings}
Note that a simple regression of time on treatment and score assumes the returns to additional time spent are the same between the treatment and control groups, which is unlikely true. Therefore, to compare speed holding accuracy constant, we employ a linear regression framework that proceeds in three steps. First, we estimate the task duration as a function of accuracy for the control group. Then we predict the task duration the control group would need to achieve the same accuracy as the treatment group. Finally, we compute the difference between task durations for the two groups, using a bootstrap to compute the level of certainty of this finding.

Let $T_g$ and $A_g$ represent the set of task durations and accuracy scores for each subject in group $g\in\{(t)reatment,(c)ontrol\}$. And let $\bar{T}_g$ and $\bar{A}_g$ represent the sample means. We then estimate the effect of accuracy on task duration for the control group via the following regression and ordinary least squares. 
$$T_c = \alpha +\beta A_c + \epsilon$$
With $\hat{\alpha}$ and $\hat{\beta}$, we next solve for the time it would take the control group to achieve $\bar{A}_t$, and call this $\Tilde{T}_c \equiv \hat{\alpha}+\hat{\beta}\bar{A}_t$. Let the difference $\Tilde{T}_c - \bar{T}_t$ be the effect of Copilot on task duration holding accuracy constant. Then, to perform inference, we simply bootstrap this procedure by sampling with replacement the control subjects and calculating $\Tilde{T}_c - \bar{T}_t$ for each sample to get our bootstrap distribution of the effect of Copilot on task duration holding accuracy constant.

\end{document}